\documentclass[12pt]{article}

\usepackage[utf8]{inputenc}
\usepackage{t1enc} 
\usepackage{graphicx}   % need for figures
\usepackage{amsmath}
\usepackage{multirow}
\usepackage{xcolor}

\usepackage[a4paper, total={6in, 8in}]{geometry}

\begin{document}

\title{Nonresonant amplification of coherent spin waves through voltage-induced interface magnetoelectric effect and spin-transfer torque}

\author{Piotr Graczyk  \\
graczyk@ifmpan.poznan.pl \\
	Institute of Molecular Physics, Polish Academy of Sciences, \\ M. Smoluchowskiego 17, 60-179 Poznan, Poland  \\
	\and 
	Maciej Krawczyk \\
	krawczyk@amu.edu.pl \\
	Faculty of Physics, Adam Mickiewicz University in Poznan, \\ Uniwersytetu Poznańskiego 2, 61-614 Poznan, Poland
	}

\maketitle

\begin{abstract}

We present new mechanism for manipulation of the spin-wave amplitude through the use of the dynamic charge-mediated magnetoelectric effect in ultrathin multilayers composed of dielectric thin-film capacitors separated by a ferromagnetic bilayer. Propagating spin waves can be amplified and attenuated with rising and decreasing slopes of the oscillating voltage, respectively, locally applied to the sample. The way the spin accumulation is generated makes the interaction of the spin-transfer torque with the magnetization dynamics mode-selective and restricted to some range of spin-wave frequencies, which is in contrary to known types of the spin-transfer torque effects. The interfacial nature of spin-dependent screening allows to reduce the thickness of the fixed magnetization layer to a few nanometers, thus the proposed effect significantly contributes toward realization of the magnonic devices and also miniaturization of the spintronic devices.

\end{abstract}

\section{Introduction}

Spin waves are potential candidates to replace electrons in logical systems with faster, ultralow energy consumption and operating at sub-micrometer scale. Much effort has been devoted so far to develop methods of spin wave generation, steering and conversion which led to the first laboratory realization of the spin-wave-based logic units \cite{Chumak2015}. However, the main challenge for the magnon computing which is relatively high attenuation, remains unresolved. Although, spin waves have usually higher propagation lengths than spin currents, still they are far behind electric signals.

We can split the methods of control of spin wave attenuation into two categories. The first relies on resonant or parametric amplification by means of microwaves \cite{Smith2007, Bracher2011, Bracher2014, Bracher2014a} or magnetoelastic \cite{Balinskiy2018, Khitun2009, Cherepov2014, Chowdhury2015,Graczyk2017b} fields and thus it enhances the signal of particular frequency. The second method relies on the exchange interaction between spin-polarized conductive electrons and localized magnetic moments through spin-transfer torque (STT). For this method the spin accumulation has to be generated in the system by means of spin-dependent conductivity \cite{Nikitchenko2019,Seo2009}, spin-dependent surface screening \cite{Graczyk2019} or various other effects related to the spin-orbit coupling: spin Hall effect \cite{An2014, Duan2014, Hamadeh2014, Padron-Hernandez2011, Wang2011}, Rashba-Edelstein effect \cite{Jungfleisch2016}, generalized spin-orbit torques \cite{Amin2016, Amin2018, Demidov2012} and topological effects \cite{Fan2016, Navabi2019}. 

STT is regarded to be non-selective, i.e., it affects all the spin-wave modes in the system regardless of the frequency or the wavevector \cite{Bracher2014a}. It means, that the transmitted signal is enhanced or suppressed together with the incoherent noise. While it is correct statement for the spin current generated by the spin Hall effects, where the polarization of nonequilibrium spins is determined by the direction of the electric field in heavy metal and geometry, but not the direction of magnetic moment in the adjacent ferromagnet. It does not necessarily apply to other mechanisms of  spin current generation. Here, we propose another approach and demonstrate, that the spin accumulation driven by the spin-dependent surface screening in ultrathin magnetoelectric laminate works differently: it effectively amplifies and modulates only coherent spin wave modes of specific symmetry and group velocity. 

In our proposition, the modulation of spin wave amplitude takes place in the magnetoelectric cell (MEC, Fig. 1) which consists of two high-permittivity TiO$_2$ nanocapacitors in series connected through conductive ferromagnetic bilayer. The thin ferromagnetic layers, made from Fe and Co, are separated by nonmagnetic metal (Cu). 

\begin{figure}
\includegraphics*[width=3in]{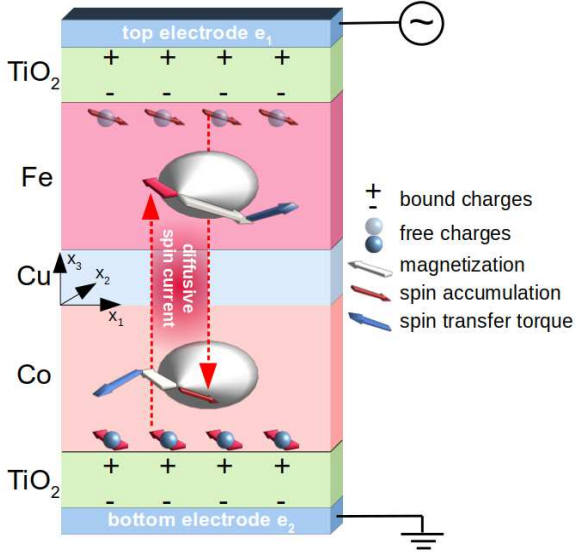}
 \caption{\label{Fig1} The principle of operation of magnetoelectric cell subjected to the ac voltage. The spin-dependent surface screening at the Fe/TiO$_2$ and Co/TiO$_2$ interfaces generates the spin accumulation which diffuses through the Fe/Cu/Co layers and exerts the spin-transfer torque on the local magnetization.}
\end{figure}

The principle of operation is as follows (see, Fig. 1). The ac voltage applied to the MEC generates time-varying screening charges at the ferromagnetic interfaces with dielectrics. Because the density of the screening charge is spin-dependent \cite{Zhang1999, Zhuravlev2010, Niranjan2009}, the dynamic spin-dependent potential produces nonequilibrium spin density at the interfaces, as we have demonstrated in Ref. \cite{Graczyk2019}. Importantly, the strength of the effect depends on the magnetoelectric constant $\gamma_s$. The spin accumulation is polarized along the local, precessing magnetic moment at the interface with dielectric, and it diffuses between the layers. If the magnetization of the second layer is non-collinear to that spin accumulation (i.e., to the magnetic moment at the surface of the first layer), it exerts a spin-transfer torque. From this, it arises that the spin-transfer torque depends on the profile of the dynamic magnetization through the bilayer thickness, i.e., the spin wave mode profile.

To describe the coupled charge-spin-magnetization dynamics of the system and demonstrate usefulness of the MEC we solve numerically the set of differential equations for the bound ($n_b$) and free ($n_f$) charge density, spin accumulation $\vec{s}$ and corresponding currents: displacement ($J_d$), charge ($J$) and spin ($J_s$) currents, electric potential $V$ and magnetization $\vec{ m}$, presented in the Methods. Although we implemented both spin-dependent surface screening and spin-dependent conductivity as a sources of the spin accumulation in the model, the latter is neglible since interface effects play dominant role in the considered system. For the interaction between spin accumulation and magnetization at interfaces we applied the continuous approach for the spin-transfer torque \cite{Lepadatu2017, Petitjean2012}.

In the following part of the paper, we briefly discuss the dispersion and phase profiles of the spin-wave modes in a dipolarly coupled ferromagnetic Fe/Co bilayers which we use in the magnetoelectric cell. We consider magnetization dynamics in the Damon-Eshbach geometry, i.e., the saturation magnetization is perpendicular to the wave vector $k$ and lies in plane of the films. Next, the magnetization dynamics for the case of uniform magnetization precession (wave vector $k = 0$) is presented with the ac voltage applied to the MEC. Then, the practical realization of the modulation and amplification of the propagating spin wave is proposed and the simulation results for the optimized MEC geometry presented. In the discussion section we analyze the symmetry, material and geometric requirements to get the effective spin wave modulation in thin film with MEC.

\section{Results}

In all the results shown below, unless otherwise stated, we applied ac voltage of amplitude $U = 10$ V and frequency $f_U = 200$ MHz to the MEC electrodes. This gives an ac charge current of the maximum density $J=10^5$ A/cm$^2$. The thickness of each layer of the TiO$_2$/Fe/Cu/Co/TiO$_2$ stack is assumed to be 2 nm, except Cu which is 5 nm thick. The physical properties of the layers are given in section \textit{Materials}.

\subsection{Spin-wave dispersion relation of the bilayer system}

The calculated dispersion relation of the two spin-wave modes of the lowest frequency in the Fe/Cu/Co system is shown in Fig. 2. The low-frequency mode (red line) with the amplitude concentrated mainly in Co is an antisymmetric mode (AS), i.e., the phase of magnetization precession between layers is shifted by $\pi$ (see insets in Fig. 2). The second mode of the higher frequency (blue line) with the amplitude concentrated mainly in Fe is a symmetric mode (S), i.e., with the in-phase magnetization precession in Co and Fe. Both modes have almost completely uniform amplitude through each layer thickness. Note, that for $k = 0$ the modes in Fe and Co are independent since there is no dipolar coupling between the layers in this case \cite{Graczyk2018a}. The higher-order modes are in the teraherz range.

\begin{figure}
\includegraphics*{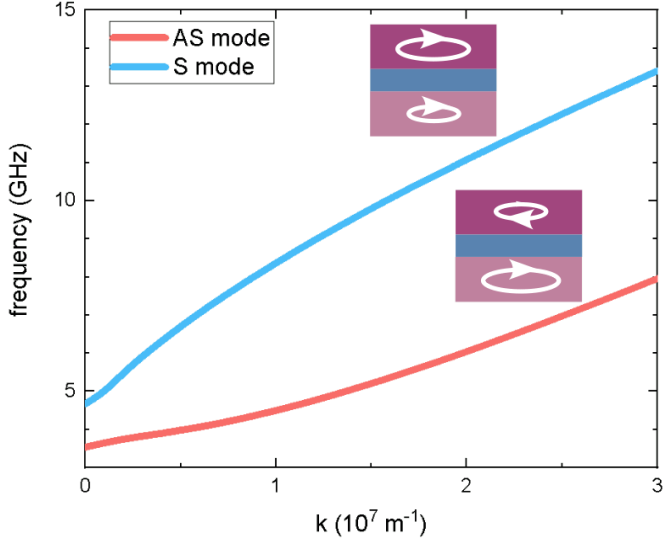}
 \caption{\label{Fig2} The spin-wave dispersion relation for the Fe/Cu/Co multilayer in the Damon-Eshbach configuration. The insets show relative phase and amplitude of the magnetization precession for the antisymmetric (AS) and symmetric (S) modes in Fe and Co layers.}
\end{figure}

\subsection{Spin wave amplification}

We will first demonstrate the effect of the ac voltage on the uniform precession of the magnetization ($k = 0$). The magnetization at the starting time ($t = 0$) is deviated from its equilibrium direction $x_2$ in the Co layer by taking initial value of the dynamic magnetization $m_1 = 10$ A/m. Then, the magnetization precesses freely with the frequency 4.3 GHz, i.e., frequency of the antisymmetric mode. If we do not apply any voltage, the precession amplitude decays exponentially with time due to the Gilbert damping as shown with black lines in Fig. 3. 

\begin{figure}
\includegraphics*[width=2.5in]{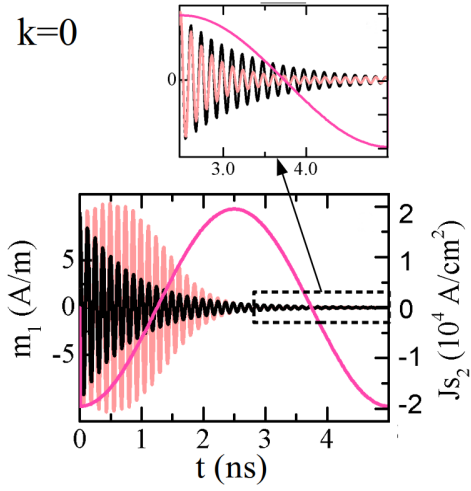}
 \caption{\label{Fig3} The time-dependence of the magnetization component $m_1$ (red line) and spin current $J_{s2}$ (magenta line) in Co in the MEC subjected to the ac voltage; black line shows $m_1$ dynamics in the absence of ac voltage.}
\end{figure}

With the ac voltage applied to the MEC, the precession of the magnetization in Co is enhanced in the first half of the voltage and current cycle, but it is strongly damped in the second half of this cycle (see, the red lines in Fig. 3). After one cycle of the voltage (5 ns) the amplitude of precession is exactly the same as in the case without the voltage. Therefore, the average impact of the voltage on the spin wave is zero. Since the Fe and Co layers are decoupled for $k = 0$ and they have different ferromagnetic resonance (FMR) frequencies, the STT does not induce coherent precession in Fe layer.

With the propagating spin wave we can map the temporal modulation of the magnetization dynamics induced by the ac voltage into the spatiotemporal distribution of the spin wave amplitude. For this purpose the MEC of finite length is chosen with the width $w_{\text{MEC}}=v/f_U$, estimated with respect to spin wave group velocity $v$ and the voltage frequency $f_U$. For that width, the wavefront entering MEC area at the time when the voltage starts to enhance the magnetization dynamics, leaves MEC area when the voltage stops to enhance, i.e., after half of the voltage period. Therefore, for this MEC width the spin wave should be maximally enhanced. 

\begin{figure}
\includegraphics*{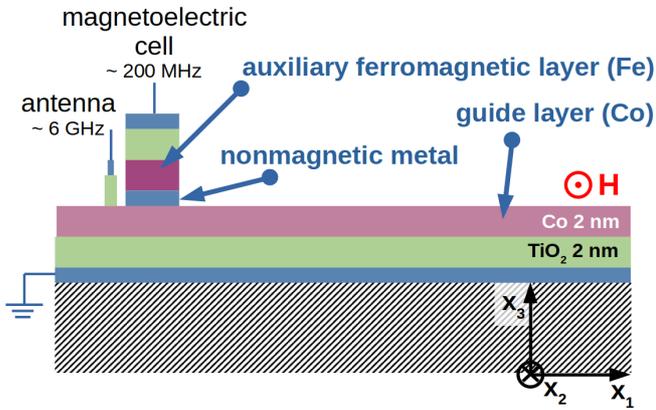}
 \caption{\label{Fig4} The spin-wave waveguide from Co thin film with the magnetoelectric cell for amplification and amplitude modulation of the spin waves. }
\end{figure}

The system proposed for the spin-wave modulation and amplification is shown in Fig. 4. It consists of the extended Co thin film and the rest of the MEC of the finite width. Because of the mode symmetry which translates into the STT direction (discussed in the next section), only the asymmetric mode (related to the Co layer) is affected by STT. Therefore, we have chosen Co layer as a guide layer for the spin wave, while Fe acts as an auxiliary layer in MEC. The width of MEC $w_{\text{MEC}} = 2.25$ $\mu$m was optimized for the group velocity of the asymmetric spin-wave mode at the frequency of 6 GHz.

\begin{figure}
\includegraphics*{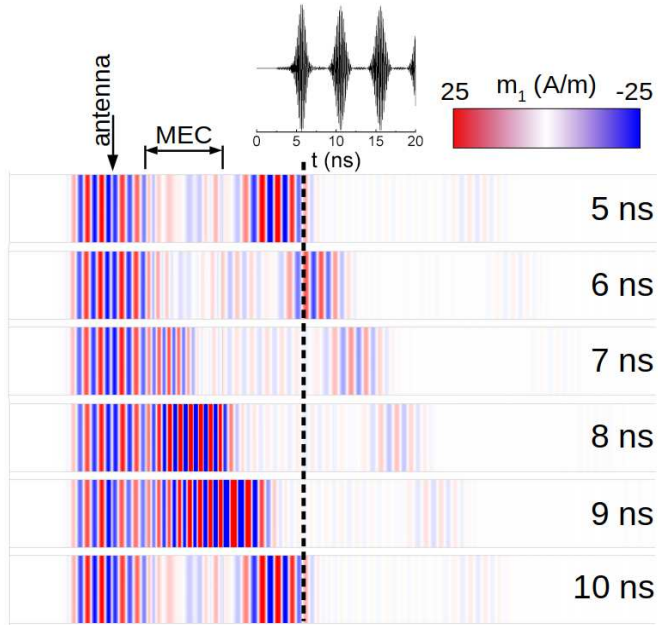}
 \caption{\label{Fig5} Magnetization component $m_1$ of the spin wave AS mode at 6 GHz in Co which passes MEC. The inset shows the time dependence of $m_1$ at the fixed point in Co film after the output of MEC. }
\end{figure}

Figure 5 shows the results of the numerical time-dependent simulations. The plane spin wave at frequency 6 GHz is continuously excited by the antenna and the wave propagates to the right entering MEC. Under the MEC the spin wave is subjected to the spin-transfer torque from the spin accumulation generated by dynamic spin-dependent surface screening. At the output we obtain spin wave with modulated amplitude. Fig. 5b shows the spin wave dynamic component $m_1$ at a fixed point behind MEC in dependence on time. The signal is periodically modulated: amplified and suppressed with the frequency of the ac voltage $f_U$  applied to MEC. When comparing the maximum amplitude of the spin wave at the output of MEC with the amplitude of the spin wave at the same position but in the absence of the ac voltage, we found the gain of the spin-wave amplitude reaching 24.

\begin{figure}
\includegraphics*{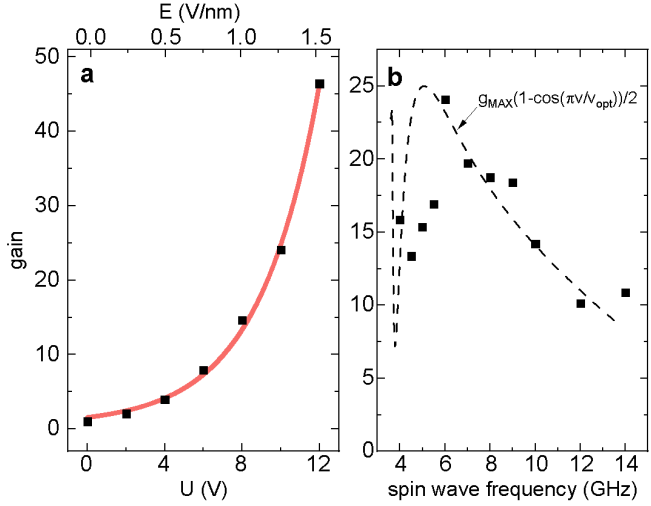}
 \caption{\label{Fig6} (a) The dependence of the maximum gain of the spin wave amplitude on the amplitude of the ac voltage in MEC; (b) the dependence of the gain on the spin wave frequency passing MEC. }
\end{figure}

Interestingly, the gain introduced by the MEC increases exponentially with the amplitude of the voltage as is shown in Fig. 6a. Further, we have also checked how the MEC of the fixed width and for the fixed voltage frequency influences the spin waves at different frequencies (Fig. 6b). As the MEC width and the voltage frequency are optimized for the group velocity of the spin wave at 6 GHz, the gain decreases for other frequencies. However, since the group velocity does not vary severely in this range of frequency, the gain remains at significant level, it is above 15 in a broad range of frequencies exceeding 5 GHz. The points achieved from the simulation in Fig. 6b are compared to the dependence calculated based on the spin-wave group velocity $v$ obtained from the dispersion relation. Here, $v_{\text{opt}}$ is the group velocity for the spin wave of 6 GHz and $g_{\text{MAX}}=24$ is the maximum gain obtained in simulations. Going from high frequencies, the gain reaches maximum at 6 GHz (for $v_{\text{opt}}$), suddenly drops at lower frequencies and increases again at the lowest frequencies. This is because the group velocity for the antisymmetric mode (comp. red line in Fig. 2) increases for $k \approx 0$.  While the theoretical line fits well for higher frequencies, some differencies are visible for lower frequencies, which is the result of the reflections at the MEC edges. 

\section{Discussion}

To understand obtained results we will describe the influence of the voltage-driven STT onto magnetization dynamics in bilayer system for the propagating spin wave. Fig. 7 shows the precession cone of the magnetization in Fe (left) and Co (right) layers for the mode S (up) and mode AS (bottom) and the magnetization direction at a given time (black arrows). In the further part, we will refer to the situation at a particular moment of the voltage/current dynamics.  As described in the introduction, the dynamic spin-dependent charge screening produces spin accumulation at the ferromagnetic-dielectric interfaces. The spin accumulation is polarized along the local, precessing magnetic moment at the interface, i.e., $\vec{s}_{\text{{Fe}}} \propto \vec{m}_{\text{{Fe}}}$ and $\vec{s}_{\text{{Co}}} \propto \vec{m}_{\text{{Co}}}$. The spin accumulation gradient $\delta \vec{s} = \vec{s}_{Fe}-\vec{s}_{Co}$ is quickly compensated as a result of the diffusive transport between layers. The resultant spin accumulation $\vec{s}=(\vec{s}_{\text{{Fe}}}+\vec{s}_{\text{{Co}}})/2$  shown in Fig. 7 by red arrows decays now further due to the STT interaction with magnetization and the spin-flip relaxation. 

\begin{figure}
\includegraphics*{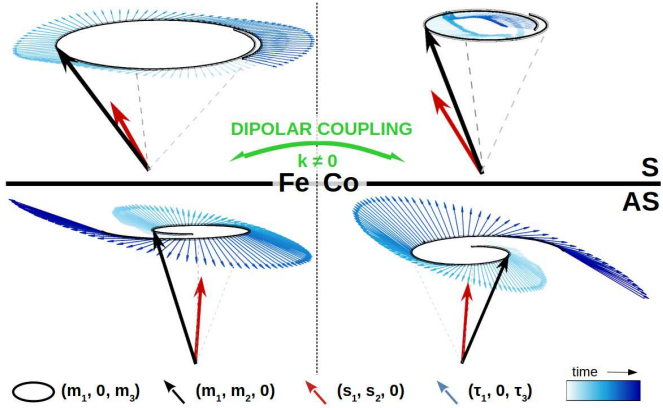}
 \caption{\label{Fig7} The precession cone of the magnetization in Fe (left) and Co (right) layer for the symmteric mode S (up) and antisymmetric mode AS (bottom), the spin accumulation and generated spin-transfer torques. The precession trajectory and spin-transfer torques are taken from numerical simulations. The spin transfer torques are averaged through each layer thickness. }
\end{figure}

Because the precession of the magnetization has different amplitude in Co and Fe layers for the mode S (upper part of Fig. 7), the spin accumulation $\vec{s}$ points inside the precession cone in Fe and it points outside the precession cone in the Co. As a consequence, the spin-transfer torque acts in the opposite direction in the both layers. Since the magnetization dynamics is coupled by the dipolar fields, the overall influence of the STT on the magnetization dynamics on the mode S is suppressed.

The bottom part of Fig. 7 shows magnetization precession for the AS mode. The spin accumulation direction points in a manner similar to the case above with respect to magnetization direction at a given time. The STT directions are also opposite between layers. However, as the precession in the layers are out-of-phase (the $\pi$ phase shift) in the AS mode, the STT direction points at outside the precession cone in both layers. Therefore, the magnetization precession is enhanced both in Fe and Co. The dipolar fields acting between layers lead to the enhancement of the AS mode. In the opposite phase of the voltage change the mode is then effectively damped, since the sign of the spin accumulation and STT is the opposite.

Now we can analise the case for or $k = 0$ more deeply. There, STT modulates the amplitude of the magnetization dynamics for both modes: FMR in the Fe layer and FMR in the Co layer. This is because for $k = 0$ the layers do not interact via dipolar fields, so the dynamics in the Fe and Co layers is coupled solely by STT. STT acts for the FMR mode in Fe in the same way as on the S mode shown in Fig. 7, i.e., it enhances precession in one layer and it reduces precession in the other layer. However, the reduction of the precession in one layer does not influence the enhancement in the other layer, because the dynamics is dipolarly decoupled. For the FMR mode in Co the STT acts to enhance the precession in both layers. However, the enhancement of precession in Fe does not strengthen precession in Co through dipolar field, so the amplification of the FMR mode is much weaker as compared to the AS mode at $k \neq 0$. Moreover, the dynamics driven by the STT onto Fe is not coherent, since the resonance frequencies of the layers are different.

To maximize the effect of the voltage-driven amplification, we have chosen ferromagnetic materials with the opposite sign of the magnetoelectric constant ($\gamma_s = -0.25$ for Co and $\gamma_s = 0.36$ for Fe). With this selection, for the opposite charge densities which accumulate at the TiO$_2$/Fe and Co/TiO$_2$ interfaces, the induced spin accumulation is parallel (or antiparallel) to the magnetization in both layers. The same sign of the spin accumulation component along the direction of the magnetization saturation ($x_2$) in the layers blocks its depletion due to the diffusion between layers (small spin accumulation gradient $\delta \vec{s}$) and thus it assures its high value in the system which is essential for high value of STT. Alternatively, one can use the bilayer of the same material but with the antiparallel magnetization alignment \cite{Graczyk2018}. In this case, both modes that exists in the system are alternated by STT. However, STT acts congruently between the layers only at sectional magnetization positions in the precession cone for the antiparallel configuration. Therefore, the modulation effect is much weaker.

In summary, we have shown numerically that the voltage-driven spin accumulation in magnetoelectric cell exerts the spin-transfer torque on the magnetization in a bilayer system. The induced charge current in the MEC is of the order of 10$^5$ A/cm$^2$. That is an order of magnitude less than the current needed for control of the attenuation by SHE shown by An et al. \cite{An2014} and 4 orders of magnitude less than control of the attenuation by spin-dependent conductivity with in-plane current shown by Seo et al. \cite{Seo2009}. Depending on the mode symmetry, STT driven by spin-dependent surface screeining may lead or may not lead to the modulation of the spin-wave amplitude.  The investigated magnetoelectric cell, although its width is optimized for the particular spin-wave velocity and voltage frequency, is able to modulate spin waves in relatively broad range of frequencies from FMR to several GHz. However, it will not have, contrary to other reported amplification techniques based on spin-orbit torques, the influence on the incoherent noise. The generation of nonequilibrium spin density through dynamic spin-dependent surface screening in the proposed magnetoelectric heterostructure allows to reduce the thickness of fixed magnetization layer used in conventional spin valve to a few nanometers, thus the proposed effect can significantly contribute to miniaturization of the spintronic devices.

\section{Methods}
\subsection{Drift-diffusion model}
The coupled charge-spin-magnetization dynamics driven by the ac voltage is considered on the base of the diffusive model \cite{Valet1993,Zhu2008, Lepadatu2017,Petitjean2012}. The free charge density current $J_f$ is described in the metal layers by the equation:
\begin{equation}
\vec {J}_f=\sigma \vec E - D \nabla n_f +\beta D \frac{e}{\mu_B} (\nabla \vec s) \hat m,
\label{eq1}
\end{equation}
which includes contributions from electron drift driven by electric field $\vec E$, diffusion driven by the gradient of free charge density $n_f$ and diffusive spin polarisation representing spin conductivity contribution, respectively. Here, $\sigma$ is conductivity, $D$ is diffusion constant, $\beta$ is spin asymmetry coefficient, $\mu_B$ is Bohr magneton and $e$ is charge of electron. Versor $\hat m$ points along magnetization direction and in the linear regime it is $\hat m= \vec m /M_s = (m_1/M_s, 1, m_3/M_s)$.

In dielectric layers the displacement current is described by the equation:
\begin{equation}
\vec{J}_b=\epsilon_0 \epsilon_r \frac{\partial \vec E}{\partial t},
\label{eq2}
\end{equation}
where $\epsilon_0\epsilon_r$ is permittivity of the material.  The conservation of free ($i=f$) and bound ($i=b$) charge density $n_i$ is described by the continuity equation:
\begin{equation}
\frac{\partial n_i}{\partial t}=-\nabla \cdot \vec{J}_i.
\label{eq3}
\end{equation}

Electric potential $V$ is given by the Gauss Law: 
\begin{equation}
\begin{split}
\epsilon_0 \Delta V
&=n_f+n_b, \\
\vec E&=-\nabla V,
\label{eq4}
\end{split}
\end{equation}
with the boundary conditions at the electrodes:
\begin{equation}
\begin{split}
V|_{x=e1} &=U\sin (2 \pi f_U t), \\
V|_{x=e2} &=0,
\label{eq5}
\end{split}
\end{equation}
where $f_U$ is the frequency of the oscillating voltage $U$.

The spin current described by the second rank tensor $\mathbf{J_s}$ in the ferromagnetic material is modeled by the equation \cite{Zhu2008}:
\begin{equation}
\begin{split}
\mathbf{J_s}=&- D \nabla \vec s -\beta\frac{\mu_B}{e}\left(\sigma \vec E - D \nabla n_f\right) \otimes \hat m \\
&- \frac{\tau}{\tau_J}\mathbf{J_s} \times \hat m - \frac{\tau}{\tau_{\perp}} \hat m \times \mathbf{J_s} \times \hat m,
\label{eq6}
\end{split}
\end{equation}
which describes spin current driven by gradient of the spin accumulation $\vec s$ (diffusion), the spin-conductivity term. The last two terms (see the description of the constants below) are the ballistic corrections to the spin-transfer torque in the diffusive model \cite{Petitjean2012}.

The continuity equation for the spin accumulation $s$ is:
\begin{equation}
\frac{\partial \vec s}{\partial t}=-\nabla \cdot \mathbf{J_s}-\frac{\vec s}{T_1}+\vec{\tau}_{\text{{STT}}}+ \vec{f}_{S},
\label{eq7}
\end{equation}
which describes the rate of change of $\vec s$ due to gradient of the spin current, the spin-flip relaxation with the characteristic time $T_1$. The spin-transfer torque $\vec{\tau}_{\text{{STT}}}$ is given by:
\begin{equation}
\vec{\tau}_{\text{{STT}}}=\tau_J^{-1} \hat m \times \vec s + \tau_{\perp}^{-1} \hat m \times \hat m \times \vec s,
\label{eq8}
\end{equation}
and it is characterized by spin precession length $\lambda_J=2\pi \sqrt{3} v_F \tau_J$ and spin coherence length $\lambda_{\perp}=\sqrt{3} v_F \tau_{\perp}$ where $v_F=1570$ nm/ps is the velocity of an electron at the Fermi level. The source term $\vec{f}_S$ from spin-dependent surface screening is \cite{Graczyk2019}: 
\begin{equation}
\vec{f}_S=\gamma_S \frac{\mu_B}{e} \frac{\partial n_f}{\partial t} \hat m.
\label{eq9}
\end{equation}

The magnetization dynamics is given by the Landau-Lifshitz-Gilbert equation with spin-transfer torque term:
\begin{equation}
\frac{\partial \vec m}{\partial t}=-\gamma \mu_0 \vec m \times \vec{H}_{eff}+\frac{\alpha}{M_s} \vec m \times \frac{\partial \vec m}{\partial t} - \vec{\tau}_{STT},
\label{eq10}
\end{equation}
where $\gamma$ is gyromagnetic constant, $\mu_0$ is magnetic susceptibility and $\alpha$ is Gilbert damping constant.

The effective field $\vec{H}_{eff}$ contains exchange, dipolar and Zeeman interactions:
\begin{equation}
\vec{H}_{eff}= \frac{2A}{\mu_0 M_s^2} \Delta \vec m - \nabla \phi + H_0 \hat{x}_2
\label{eq11}
\end{equation}
where $H_0$ is external magnetic field applied along direction $\hat{x}_2$. The magnetic potential $\phi$ is given from the Maxwell equation:
\begin{equation}
\Delta \phi = \nabla \cdot \vec m.
\label{eq12}
\end{equation}

Equations (\ref{eq1})-(\ref{eq12}) are solved numerically by finite element method in Comsol Multiphysics with time-varying voltage applied to the MEC electrodes. The equations are implemented with the currents defined as a flux, thus assuring their continuity at interfaces and zero value at outer boundaries. 

\subsection{Materials}
The relative permittivity is $\epsilon_r=120$ for TiO$_2$ and $\epsilon_r=1$ for metals. We assume the conductivity of metals $\sigma=1.2 \cdot 10^7$ S/m and the diffusion constant $D=4 \cdot 10^{-3}$  m$^2$/s. The spin asymmetry coefficient is $\beta=0.5$ \cite{Zhu2008}, the spin-flip relaxation time is $T_1=0.9$ ps \cite{Zhu2008}. This corresponds to the spin relaxation length of 60 nm, i.e., much larger than electron mean free path, satisfying diffusive limit \cite{Valet1993, Penn2005, Gmitra2009}. The characteristic times for the spin-transfer torques are $\tau_J=0.3$ fs, $\tau_{\perp}=1.2$ fs and $\tau_J=0.2$ fs, $\tau_{\perp}=4.4$ fs for Co and Fe, respectively \cite{Petitjean2012}. The magnetoelectric coefficient calculated from Eq. (\ref{eq10}) in Ref. \cite{Graczyk2019} is $\gamma_S=-0.25$ for Co and $\gamma_S=0.36$ for Fe. The exchange constants were taken $A=20$ pJ/m both for Co and Fe, Gilbert damping $\alpha_{\text{{Co}}}=0.005$, $\alpha_{\text{{Fe}}}=0.002$ and the saturation magnetizations $M_{\text{{s,Co}}}=1$ MA/m and $M_{\text{{s,Fe}}}=1.7$ MA/m. The external magnetic field straength is $H_0=10$ kA/m.

\section*{Acknowledgement}
The study has received financial support from the National Science Centre of Poland under grant 2018/28/C/ST3/00052.

\bibliographystyle{ieeetr}

\end{document}